\documentstyle[12pt]{article}
\newcommand{\bb}{\begin {minipage} {3cm}\begin{center}}
\newcommand{\ee}{\end{center}\end{minipage}}
\newcommand{\bc}{\begin {minipage} {2.5cm}\begin{center}}
\newcommand{\bd}{\end{center}\end{minipage}}
\newcommand {\f}{\footnote}
\newcommand {\q}{\begin{quote} \small}
\newcommand {\be}{\begin {eqnarray}}
\newcommand {\e}{\end {eqnarray}}

\renewcommand{\baselinestretch}{2}
\begin{document}
\newtheorem {lemma}{Lemma}
\newtheorem {theorem}{Theorem}
\newtheorem {coro}{Corollary}
\vspace {1cm}
\begin {center}
{\Huge Unified Treatment of }\\
{\Huge EPR and Bell Arguments in }\\
{\Huge Algebraic
Quantum Field Theory}
\\
\vspace{2.5cm}
{\em\Large Michael L.G. Redhead}\\
\vspace{0.2cm}
{\em Department of
History and Philosophy of Science,\\
Free School Lane, Cambridge CB2 3RH}\\
\vspace{1cm}
{\em \Large Fabian Wagner}\\
\vspace{0.2cm}
{\em Department of Applied Mathematics and Theoretical Physics,\\
Silver Street, Cambridge CB2 9EW}\\
\vspace{2cm}
\end{center}
\vspace{1cm}\\
\renewcommand{\baselinestretch}{1}
{\bf Abstract.} A conjecture concerning vacuum correlations in axiomatic
quantum field theory is proved. It is shown that this result can be
applied both in the context of EPR-type experiments and Bell-type experiments.
\pagebreak
\section{Introduction}
\normalsize
In the past, two arguments have 
played major roles in the discussion of the foundations and interpretations
of quantum theory. On the one hand there is the incompleteness argument by
Einstein, Podolsky, and Rosen (EPR)\cite {epr}. For EPR a necessary condition
for a theory to be complete is that `every element of the physical reality
must have a counterpart in the physical theory' (\cite{epr}, 777). They consider
a setup, for which, they argue, one can find an element of reality that is not
represented by anything in the formalism of quantum mechanics. In Bohm's version
of the argument two spin-$\frac{1}{2}$ particles move in opposite directions
as a result of a decay or scattering process. The system is described by the singlet
state in which the spins of the two particles are perfectly correlated, that is,
by measuring the spin of one of the particles in an arbitrary direction the 
outcome of the same measurement on the other particle can be predicted with 
probability one. Assuming further that elements of reality relevant to one
part of the system cannot be affected by measurements performed at spacelike 
distance on another part of the system, EPR show that an element of reality
at the unmeasured particle must have existed {\em before} the measurement
on the spacelike separated particle. Since this element of reality has no 
counterpart in the formalism of quantum mechanics, one can conclude that the 
quantum mechanical description is incomplete. This `streamlined'
version of the EPR argument is set out in more detail in \cite{red4}. In effect the realist
can use this argument to refute not only the instrumentalist interpretation but also
Bohr's complementarity interpretation, by assuming a suitably specialised locality
condition. 

On the other hand, the realist himself runs into difficulties (and this is
the second argument) if he assumes a more complete description in terms
of hidden variables and, in addition, the appropriate locality condition
that any sharp value for an observable cannot be changed into another sharp 
value by altering the setting of a spacelike separated piece of apparatus \cite{red4}. The realist then 
faces the problem that the Bell inequality derived under these assumptions
is in general violated. Thus, unless one gives up the locality condition,
the realist hidden variable interpretation is not tenable.

In algebraic quantum field theory, which provides a relativistic framework of quantum theory,
aspects of these arguments can be generalised. It was shown by one of
us (MLGR)
that perfect correlations, as they appear in Bohm's version of the
EPR-argument, can effectively be demonstrated in quantum field theory, 
that is, by appropriate choice of (idealised) measurement
devices perfect correlations can be approximated arbitrarily
closely - even in the vacuum state \cite {red3}. On the other hand it was shown by Landau,
Summers and Werner that Bell inequalities are violated in the vacuum state,
and that a maximal violation can be achieved by conditioning the Bell inequality
on the outcome of a third, spacelike separated measurement
\cite{landau1}\cite{landau2}\cite{s1}-\cite{summ4}. The object of the
present paper is to show, perhaps surprisingly, that these two claims
can be derived from a single `root' theorem , a proposal put forward recently by
Malament \cite{mall}, although no proof of the result has hitherto
been published. 

The paper is organised as follows. First we briefly state some axioms
in order to specify our framework. We then proceed to prove the root
theorem. The proofs of the lemmata we employed are given in the
appendix. In section \ref{rtb} we show how previous results in the
disussion of the two distinct arguments can be recovered as special cases of
the root theorem. 

\section{Axiomatic QFT}\label{alla}

In this section we shall review briefly the axiomatic framework and
the result known as the Reeh-Schlieder theorem. 
Although not all of the axioms presented here are needed to prove the Reeh-Schlieder theorem,
section \ref{rtb} requires a more specific framework for the analysis of 
Bell-type correlations which is why an extended set of axioms is
presented here \cite {butter} \cite {landau1}.

Let $\{O_i\}$ be bounded open regions in Minkowski space. Local
observables are identified with hermitean elements in local
(non-commutative) von Neumann-algebras 
${\cal R}(O_i)$ 
which are subsets of the set of bounded operators, denoted by ${\cal B(H)}$,  acting on some Hilberspace
${\cal H}$. The axioms we want to impose are then:

\renewcommand{\labelenumi}{(\roman{enumi})}
\begin {enumerate}
\item {\bf Isotony:} If $O_1 \subseteq O_2$, then ${\cal R}(O_1)\subseteq {\cal R}(O_2)$.
\item {\bf Spacelike commutativity:} If all points in $O_1$ are spacelike separated
from all points in $O_2$, then ${\cal R}(O_1) \subset \left[{\cal R}(O_2)\right]'$,
that is, every operator in  ${\cal R}(O_1)$ commutes with every operator in  ${\cal R}(O_2)$.
\item {\bf Poincar\'e Covariance:} There exists on ${\cal H}$ a strongly continuous
unitary representation $U(P)$ of the (covering group of the) Poincar\'e group 
which meshes with the automorphisms of the global algebra ${\cal R}$, in the 
following sense:
 to each (active) Poincar\'e transformation $g:=(\Lambda, a)$
acting on Minkowski spacetime, there is associated an automorphism 
$\displaystyle {\alpha}_g$ of the global algebra ${\cal R}$ with 
$\displaystyle {\alpha}_g\left[{\cal R}(O)\right]={\cal R}\left[g(O)\right]$
and such that $U(g)AU(g)^{-1} =\displaystyle {\alpha}_g(A)$ for all $A\in 
{\cal R}(O)$. 
\item {\bf Diamond:} Let $D(O)=D^+(O)\cup D^-(O)$ be the domain of dependence 
(sometimes called causal shadow) of the spacetime region $O$.\f{The future 
domain of dependence, $D^+(O)$, is defined as the set of points in spacetime
through which every past inextendible causal curve intersects $O$. For the
definition of $D^-(O)$ interchange `past' with `future'.} Then
${\cal R}(O) = {\cal R}\left[D(O)\right]$.
\item {\bf Weak additivity:} For any non-empty region $O$, the set $C''$ equals
the von Neumann algebra generated by the set of all translations of ${\cal R}(O)$,
that is, $C''=\left\{\bigcup_{a\in {\bf R}^4}{\cal R}(O_a)\right\}''$.
\item {\bf Irreducibility:} ${\cal H}$ has no non-trivial subspace invariant under
all elements of all ${\cal R}(O)$, that is, $C''={\cal B(H)}$.
\item {\bf Spectrum:} The spectrum of the translation subgroup of the Poincar\'e
group is contained in the closed forward lightcone, that is, the energy-momentum
eigenvalues are restricted by $\vec{p}^{\:2}\ge 0$, $p^0\ge 0$.
\item {\bf Vacuum:} There is a unique state $\Omega$, called the vacuum, that is 
invariant under all actions of the Poincar\'e group: 
$\forall g \in {\cal P}:
U(g)\Omega=\Omega.$

\end {enumerate}

 A
state $\psi$ in ${\cal H}$ for which the set $\left\{A\psi: A\in{\cal A}\right\}$
is dense in ${\cal H}$ is called a cyclic vector for ${\cal A}$ with
respect to ${\cal H}$.
We have now introduced all the necessary concepts in order to state without
proof 
\begin {theorem}[Reeh and Schlieder \cite{reeh}:]\label{rs}
Let $O$ be an open bounded set in spacetime. Then $\Omega$ a is cyclic vector
for ${\cal R}(O)$.\hfill$\Box$
\end{theorem}
Note that the theorem does not merely state that the state $\Omega$ is a cyclic
vector for the global algebra ${\cal R}$. The latter is certainly what we would
expect even in the Schr\"odinger approach to quantum field theory.

It is worth noting that the vacuum can be replaced by any state with bounded 
energy in the Reeh-Schlieder theorem. Furthermore, the theorem may be
further strengthened by replacing `open bounded set' by `open set with
non-empty causal complement'. Theorem \ref{rs} has an important 
consequence, expressed in the following
(theorem 5.3.2 in Haag \cite{haag})
\begin {theorem}\label{sep}
If $O$ has non-void causal complement, then ${\cal R}(O)$ does not contain any
operator which annihilates the vacuum, that is, if $A\Omega=0$ for some $A
\in {\cal R}(O)$, then it follows that $A=0$. The vacuum is therefore called
a {\em separating vector}.\hfill$\Box$
\end{theorem}

This property ensures that, if two operators $A_1$ and $A_2$ in ${\cal R}(O)$
generate the same state from the vacuum, then they have to be the same operator.
It also shows that there are no pure local annihilation operator
and thus, using the fact that ${\cal R}(O)$ is an involutive algebra, no pure
localised creation operator. This supports the view that the concept of a quantum 
is a non-local one.

Theorem \ref{sep} also implies the (Theorem 2 in Redhead \cite{red3}, 
Corollary in Hellweg and Kraus \cite{hkr})
\begin {coro}\label{co1}
Any possible outcome of any possible measurement procedure will occur with 
non-vanishing probability in the vacuum.\hfill$\lhd$
\end {coro}

\section {The root theorem}\label{mal}
The Reeh-Schlieder theorem provides us with a powerful tool in the analysis of 
non-locality. It shows
that localised causes can have non-local effects, even in the vacuum state. The
following theorem determines more specifically the kind of effects that can be 
produced by selective local actions on the vacuum state. In particular, it deals
with two open bounded spacelike separated regions, in each of which measurements
can be performed.

\begin {theorem} [Root Theorem] \label{t1} For any two spacelike separated bounded open regions
$O_1$ and $O_2$, for any self-adjoint operator $A$ in ${\cal R}(O_2)$, and for any unit
vector $\psi$, if $\langle A\rangle_{\psi}\:=K$ then, for all $\varepsilon > 0$, there are 
projection operators $P_1$ and $P^{\natural}_1$ in ${\cal R}(O_1)$ such that
\be
\langle AP_1\rangle_{\Omega}\: &>&(K-\varepsilon)\langle P_1\rangle_{\Omega}, \quad \mbox{and}\label{m1}\\
\langle AP^{\natural}_1\rangle_{\Omega}\: &<& (K+\varepsilon) \langle P^{\natural}_1\rangle_{\Omega},\label{zugabe}
\e
where $\Omega$ is the vacuum state.
\end{theorem}

{\sc Remarks:} First, notice that it is not required that $\langle
A\rangle_{\psi}=K$ for all unit states; it merely has to hold for at least 
one unit state.
Secondly, as $P_1\neq P^{\natural}_1$ in general, it is not claimed
that in general $K$ can be approximated arbitrarily closely.
Third, we will assume that $A$ is defined everywhere on the Hilbert space
${\cal H}$. By the theorem of Hellinger and Toeplitz \cite{reed}, $A$ is then bounded, that is, $\|A\|$ exists and is finite. 
In particular, for any $\phi \in {\cal H}$ we have $\|A\phi\|\le\|A\| \|\phi\|$. Also,
$\Omega$ can be replaced by any state with bounded energy.\\

{\bf\sc Proof of Theorem \ref{t1}:} Let $A \in {\cal R}(O_2)$,
and let  $\psi$ be a unit vector for which 
$\langle A\rangle_{\psi}=K$. By the Reeh-Schlieder
theorem, $\psi$ can be approximated arbitrarily closely by elements of the 
set ${\cal R}_1\Omega=\left\{C\Omega: C\in {\cal R}(O_1)\right\}$, that is
\be
\forall \varepsilon_1 > 0, \exists \tilde{C}_{\varepsilon_1}\in {\cal R}(O_1), 
\mbox{such that}\quad
\|\psi- \tilde{C}_{\varepsilon_1}\Omega\|<\varepsilon_1.\nonumber
\e
It is convenient to choose  the norm of the vector
$\tilde{C}_{\varepsilon_1}\Omega$ to be equal to $1$. The following lemma ensures 
that such a choice is always possible.

\begin {lemma}\label {l1}
For all $\varepsilon_2>0$ there exists an operator  
$C_{\varepsilon_2}$ such that:
\center{$\|\psi-C_{\varepsilon_2}\Omega\|<\varepsilon_2
\quad\! \mbox{and}\!\!
\quad\|C_{\varepsilon_2}\Omega\|=1.$}\\ 
\end{lemma}

Intuitively it is clear that the expectation value of $A$ in the state 
$C_{\varepsilon_2}\Omega$ is close to $K$. This can be 
proved rigorously (Theorem 4 in \cite{licht}):

\begin {lemma}\label {l2}
For all $\varepsilon_3>0$, there exists a $C_{\varepsilon_2}$ such that
\begin{center}
$K - \varepsilon_3 < \langle A\rangle_{C_{\varepsilon_2}\Omega} < K+
\varepsilon_3.$
\end{center}
\end{lemma}\vspace{0.5cm}

 From our general assumption of
spacelike commutativity (axiom (ii)) we conclude that $A \in {\cal R}(O_2)$ 
and $C^{\dagger}_{\varepsilon_2}
\in {\cal R}(O_1)$ commute, that is,
\be
\langle A \rangle _{C_{\varepsilon_2}\Omega}=({C_{\varepsilon_2}\Omega}, A{C_{\varepsilon_2}\Omega})
= (\Omega, C^{\dagger}_{\varepsilon_2}A{C_{\varepsilon_2}\Omega}) = 
(\Omega, AC^{\dagger}_{\varepsilon_2}{C_{\varepsilon_2}\Omega}).\nonumber
\e
The operator $C^{\dagger}_{\varepsilon_2}{C_{\varepsilon_2}}\stackrel{\rm def.}{=:}Q_1$
is clearly self-adjoint and bounded, and in particular it is positive.
Hence, we may approximate it arbitrarily closely by a finite sum of projectors
with positive coefficients, that is, 
$\forall \tilde {\varepsilon}_4,  \exists n \in {\bf I\!N}$ 
and there is a family of projectors $\{P_i\}^n_{i=1}\subset {\cal R}(O_1)$ together 
with coefficients $\{\tilde{\lambda'}_i\}_{i=1}^n, 
\; \tilde{\lambda'}_i>0 \; \forall i$, such that
\be\label{qq}
\|C_{\varepsilon_2}^{\dagger}C_{\varepsilon_2}-\sum_{i=1}^n\tilde{\lambda'}_iP_i\|
\stackrel{\rm def.}{=:}\|Q_1-{\tilde{Q}_1}'\|<\tilde{\varepsilon}_4
\e
This just follows from the spectral theorem. Again, for our purposes it will
be convenient to impose the requirement that the vacuum expectation value
of the approximating operator should equal $1$. The following lemma shows 
that such a choice is possible.

\begin {lemma}\label{l3}
It is always possible to rescale the coefficients
$\tilde{\lambda}_i' \rightarrow \lambda_i'=\tilde{\lambda}_i'/
\langle \tilde{Q}'_1\rangle_{\Omega}$ such that $Q'_1=\sum\limits^n_{i=1}
\lambda_i'P_i$ is an approximation to $Q_1$ in the sense of inequality {\rm(\ref{qq})},
and  $\langle Q_1'\rangle_{\Omega}=1$.
\end{lemma}

We now incorporate the approximation to the operator $Q_1$ into lemma \ref{l2}
and find
\begin {lemma}\label{l4}
For all $\varepsilon_5>0$ there is an operator $Q_1'=\sum\limits^n_{i=1}
\lambda_i'P_i$ in ${\cal R}(O_1)$, such that
\begin{center}$ 
K-\varepsilon_5 < \langle A Q_1'\rangle_{\Omega} < K+ \varepsilon_5
\nonumber.
$\end{center}
\end{lemma}

Now we can write
\be
\langle AQ_1'\rangle_{\Omega} =\sum_{i=1}^n w_i\frac {\langle AP_i\rangle_{\Omega}}
{\langle P_i \rangle_{\Omega}}, \quad\mbox{where}\quad w_i = \lambda_i\langle
P_i\rangle_{\Omega}\nonumber
\e
and therefore $\sum\limits_{i=1}^n w_i=1$ as we have chosen using lemma \ref{l3}. Because of the 
latter property it is clear that at least one element of the set 
$\left\{\frac {\langle AP_i\rangle_{\Omega}}{\langle P_i \rangle_{\Omega}}\right\}$
is greater or equal to $\langle AQ_1'\rangle_{\Omega}$. If this was not the case, then
\be
\langle A Q_1'\rangle_{\Omega}=\sum_{i=1}^n w_i \frac {\langle AP_i\rangle_{\Omega}}
{\langle P_i \rangle_{\Omega}}< \sum_{i=1}^n w_i \langle A Q_1'\rangle_{\Omega}
=\langle A Q_1'\rangle_{\Omega}\nonumber,
\e
which is a contradiction. Therefore, 
\be
K-\varepsilon_5 < \langle AQ_1'\rangle_{\Omega} \le \quad \mbox{max}
\left\{
\frac{\langle AP_i\rangle_{\Omega}}{\langle P_i \rangle_{\Omega}}   
\right\} 
\stackrel {\rm def.}{=:}
\frac {\langle AP_{\mbox{\tiny max}}\rangle_{\Omega}}
{\langle P_{\mbox{\tiny max}} \rangle_{\Omega}}\nonumber
\e
A similar argument leads to 
\be
K+\varepsilon_5 > \langle AQ_1'\rangle_{\Omega} \ge \quad \mbox{min}
\left\{
\frac{\langle AP_i\rangle_{\Omega}}{\langle P_i \rangle_{\Omega}}   
\right\} 
\stackrel {\rm def.}{=:}
\frac {\langle AP_{\mbox{\tiny min}}\rangle_{\Omega}}
{\langle P_{\mbox{\tiny min}} \rangle_{\Omega}}\nonumber
\e
Finally, we conclude that there are projectors $P_1=P_{\mbox{\tiny max}}$ and
$P^{\natural}_1=P_{\mbox{\tiny min}}$ in ${\cal R}(O_1)$, such that 
\be\langle
AP_1\rangle_{\Omega} &>& (K-\varepsilon_5)\langle P_1\rangle_
{\Omega}, \quad\mbox {and}\nonumber\\
\langle
AP^{\natural}_1\rangle_{\Omega} &<& (K+\varepsilon_5)\langle P^{\natural}_1\rangle_
{\Omega}\nonumber
\e
which concludes the proof of theorem \ref{t1}.\hfill $\Box$\\

\section {Applications}\label{rtb}
The root theorem holds in a quite general context. The expectation value of
 any local observable $A$ belonging to a spacetime region $O_2$, in any state
$\psi$ which attains a maximum (or minimum) value, can be approximated arbitrarily closely by acting with an appropriate
projector in a spacelike separated region on the vacuum state. A special case
of such an operator $A$ is a projector, say $P_2$, attached to the region 
$O_2$. This corresponds to the situation in which a measurement device is prepared
to register a particular, single eigenvalue of an observable. Such a
 configuration has been investigated before by one of the present
 authors and formulated
as (theorem 4' in reference \cite{red3})
\begin {theorem}\label{thred}
For any two spacelike separated bounded open regions $O_1$ and $O_2$, and
for all $\varepsilon > 0$ and all projectors $P_2 \in {\cal R}(O_2)$, there
exists a projector $P_1 \in {\cal R}(O_1)$, such that
\be\label {red}
\langle P_1\rangle_{\Omega} \ge \langle P_1 P_2\rangle_{\Omega} > (1-\varepsilon)
\langle P_1\rangle_{\Omega}
\e
\end{theorem}
{\bf \sc Proof of Theorem \ref{thred}:} Let $P_2$ be a projector in ${\cal R}(O_2)$ such 
that $P_2\phi\neq0$ for some state $\phi$. Such a state can always be found: 
as $\Omega$ is a separating vector it has the required property.
Then the state $\psi=P_2\phi/\|P_2\phi\|$ has the property
$\langle P_2\rangle_{\psi}=1$. The root theorem can now be applied for $K=1$ with 
the given assignments, and so the second inequality in (\ref{red}) corresponds to
inequality (\ref{m1}), while the first one is trivial as $P_2$ is a
projector.\hfill$\lhd$\\

Theorem \ref{thred} says effectively that given a measurement outcome
of one for $P_1$, we can predict with probability as close to
unity as we like the outcome of measuring a projector at spacelike
separation. This is all that is needed to run a vacuum version of the
EPR argument.  

Let us now turn to the discussion of an application of the root
theorem to Bell inequalities.  Let $P_i$ and $Q_i$ be projection
operators in some operator algebras ${\cal A}$ and ${\cal B}$,
respectively, that act on the same Hilbert space. Let 
 $A_i=(2P_i-1)$ and $B_i=(2Q_i-1)$ be
corresponding contractions. We call the operator $R=A_1(B_1+B_2) + A_2(B_1-B_2)$ the 
{\em Bell operator} and define
\be
\beta(\phi, {\cal A}, {\cal B})=\frac{1}{2} \sup \langle R\rangle_{\phi},\nonumber
\e
the supremum over all self-adjoint contractions $A_i \in {\cal A}, B_i\in
{\cal B}$, and call $\beta (\phi, {\cal A}, {\cal B})$ the maximal Bell correlation
of the pair $({\cal A}, {\cal B})$. It is not difficult to show that Bell's
inequality is not only violated in quantum theory, but also that there exists
an upper bound for the violation, that is, a corresponding `Bell inequality'
in the quantum theory (proposition 2, Landau \cite{landau1}):
\begin {lemma}\label{bel}
For any $C^*-$algebra ${\cal C}$, any commuting non-abelian subalgebras ${\cal A}$
and ${\cal B}$ and any state $\phi$ in the dual of ${\cal C}$, 
\be
|\beta(\phi, {\cal A}, {\cal B})|\le \sqrt{2} \nonumber
\e
\end{lemma}
{\sc  Remark:} This has to be compared with the  
Bell inequality for classical observables, expressed as $\beta(\phi, {\cal
A}, {\cal B})\le 1$.\\

Note that lemma \ref{bel} does not ensure that there is a state $\phi$ 
for which the
Bell inequality is {\em maximally} violated, that is, for which
$\frac{1}{2}\langle R \rangle_{\phi} = \sqrt{2}$; nor is it clear apriori
whether, for 
a given state, there are projectors such that the violation is maximal.
However, Landau proved \cite{landau1} both propositions
for a class of von Neumann algebras, and we will formulate them as 
\begin {lemma}\label{land35}
Let ${\cal A}$ and ${\cal B}$ be two commuting non-abelian subalgebras in
a von Neumann algebra ${\cal C}$, which have the Schlieder property. Then
\renewcommand{\labelenumi}{(\roman{enumi})}
\begin {enumerate} 
\item There is a state $\tilde{\phi}$ on ${\cal C}$ such that
$|\beta(\tilde{\phi}, {\cal A}, {\cal B})|=\sqrt{2}$
\item In addition we can find contractions $A_1, A_2 \in {\cal A}$ and
$B_1, B_2 \in {\cal B}$ for which
$\frac{1}{2}\langle R\rangle _{\tilde{\phi}}=\sqrt{2}$.\hfill$\lhd$
\end{enumerate}
\end{lemma}

So far we have only considered the classical and the general quantum theoretical
situation. We now return to the more specific framework of algebraic quantum
field theory with its distinct underlying spacetime structure. After some
careful preparations, lemma \ref{land35} can be formulated for local 
algebras without difficulty.

The first thing to notice is that, according to axiom (ii), local 
algebras associated with spacelike separated regions $O_1$ and $O_2$ commute.
Thus, they are natural candidates for the algebras ${\cal A}$ and ${\cal B}$
in lemma \ref{land35}, and we expect to find maximal violation of the Bell
inequality for localised algebras. However, it can be shown that if the stronger
condition of {\em strictly} spacelike separation holds for
$O_1$ and $O_2$, then local algebras ${\cal R}(O_1)$ and ${\cal R}(O_2)$ posses
the Schlieder property. This is not true in general for spacelike
separated regions.  So we can prove 
\begin {coro}\label{c2}
Let $\left\{{\cal R}(O)\right\}$ be a net of localized von Neumann subalgebras
of ${\cal B(H)}$ subject to the axioms (i)-(viii). Then, for any strictly
spacelike separated regions $O_1$ and $O_2$ there exists a state $\phi$ on 
${\cal B(H)}$ such that $\beta(\phi, {\cal R}(O_1), {\cal R}(O_2))=\sqrt{2}$ and
there are contractions $A_1, A_2 \in {\cal R}(O_1)$ and
$B_1, B_2 \in {\cal }(O_2)$ for which
$\frac{1}{2}\langle R\rangle _{\phi}=\sqrt{2}$.
\hfill$\lhd$
\end{coro}

{\sc Remark:} Strict spacelike separation is sufficient but not
necessary for ${\cal A}(O_1)$ and ${\cal A}(O_2)$ to have the Schlieder property.
For example, any pair of spacelike separated double cone algebras
or wedge algebras have the Schlieder property as a result of the general theorem
by Driessler (Theorem 3.5. in \cite{driessler}).

The state $\phi$ for which $\beta(\phi, {\cal A}(O_1), {\cal A}(O_2))=\sqrt{2}$
holds is only one among many (typically infinitely many) states for which
the Bell inequality is violated \cite{s1}. Since the
expectation value of $R$ is a {\em continuous} functional on the Hilbert space
we can expext that there are projectors in the algebras ${\cal A}(O_1)$ and 
${\cal A}(O_2)$ and states in ${\cal H}$ which violate Bell's inequality
arbitrarily closely to $\sqrt{2}$. (Corollary \ref{c2} holds for
arbitrarily distant regions $O_1$ and $O_2$. However, for fixed state
$\phi$ the degree of violation depends on their spatial separation
\cite{fred} \cite{summ2}).
Recalling from above
that, by the Reeh-Schlieder theorem, any state in ${\cal H}$ can be approximated
arbitrarily closely by a local operation on the vacuum, we can also expect that
Bell's inequality can be violated `almost' maximally by a selective
measurement on the vacuum. This idea was developed by Landau \cite{landau2} 
and we shall now see how his result follows directly from
the root theorem.

Consider three strictly spacelike separated regions in spacetime, $O_1, O_2$
and $O_3$. The regions $O_1$ and $O_2$ contain measurement apparatuses for
measuring observables, $P_i, Q_i, i=1,2$ (the $P$'s in $O_1$ and
the $Q$'s in $O_2$), represented by operators in ${\cal A}(O_1)$ and
${\cal A}(O_2)$, respectively. As before, the self-adjoint contractions will
then take values in $[-1,+1]$ in all states. In the third region $O_3$ we consider
a single projection operator $P_3$. Following Landau we define for each
of the four joint measurements of the observables $(A_1, B_1, P_3)$,
$(A_1, B_2, P_3)$, $(A_2, B_1, P_3)$ and $(A_2, B_2, P_3)$ in the vacuum state
the {\em conditional expectation} for each of these, given the result
$P_3=1$. For example the conditional expectation for the first set of
observable is denoted by $\langle A_1 B_1\rangle_{P_3=1}$ which can be expressed
as 
\be
\langle A_1 B_1\rangle_{P_3=1}=\frac{\langle A_1B_1P_3\rangle_{\Omega}}
{\langle P_3\rangle_{\Omega}}.\nonumber
\e
The conditional expectation value of the Bell operator $R$ is accordingly
\be
\langle R\rangle_{P_3=1}=\frac{\langle A_1B_1P_3\rangle_{\Omega}}
{\langle P_3\rangle_{\Omega}} + \frac{\langle A_1B_2P_3\rangle_{\Omega}}
{\langle P_3\rangle_{\Omega}} + \frac{\langle A_2B_1P_3\rangle_{\Omega}}
{\langle P_3\rangle_{\Omega}} - \frac{\langle A_2B_2P_3\rangle_{\Omega}}
{\langle P_3\rangle_{\Omega}}\nonumber.
\e
If the theory supports a common joint conditional distribution for
$A_1$, $A_2$, $B_1$, $B_2$, then
\be
\left|\frac{1}{2}\langle R\rangle_{P_3=1}\right|\le 1 \qquad\mbox
{(conditional Bell inequality)}
\e
would hold. However, in quantum theory, the conditional Bell inequality is 
in general violated and in addition, starting from the vacuum state it can
be violated `almost' maximally, as we see from the following (proposition
1 in Landau \cite{landau2})
\begin {theorem}\label {lan3}
Let $O_1, O_2, O_3$ be strictly spacelike separated regions and $\varepsilon>0$.
Then there are self-adjoint contractions $A_1, A_2$ in ${\cal R}(O_1)$ and
$B_1, B_2$ in ${\cal R}(O_2)$, and a projector $P_3$ in ${\cal R}(O_3)$ such that
in the vacuum state
\be\label{last}
\frac{1}{2}\langle R \rangle_{P_3=1}>\sqrt{2}-\varepsilon
\e
\end{theorem}
{\sc Proof of Theorem \ref{lan3}:} Lemma \ref{land35} ensures that we can 
find a state $\tilde{\phi}$ and contractions $A_1, A_2 \in {\cal R}(O_1)$ and
$B_1, B_2 \in {\cal R}(O_2)$, such that $\frac{1}{2}\langle R\rangle_{\tilde{\phi}}
=\sqrt{2}$. Now write $\tilde{O} = O_1 \cup O_2$. Then $R \in {\cal R}(\tilde{O})$
and we can apply the root theorem for strictly spacelike separated 
$\tilde{O}$ and $O_3$: given $\frac{1}{2}\langle R\rangle_{\tilde{\phi}}
=\sqrt{2}$ theorem \ref{t1} ensures that there is a projector $P_3$ in $O_3$
such that inequality (\ref{last}) holds.\hfill$\Box$
\linebreak

{\sc Remark:} This theorem can be generalised to arbitrary observables
$A_1, A_2$ and $B_1, B_2$ that are subject to $\left[A_1, A_2\right]
\neq 0\neq \left[B_1, B_2\right]$.

\section {Conclusion}

We have seen that the vacuum has, in a sense, the property of maximal
entanglement, viz. in the sense of maximal correlation we have explained. This means that for all kinds of experiments,
performed in any bounded  region in spacetime, 
the vacuum can be collapsed into a state which gives 
a fixed extremal expectation value arbitrarily precisely. This collapse
can be achieved by a selective measurement
in a spacelike separated region. 
As a consequence we have shown that in quantum field theory perfect
correlations, such as used in the EPR-argument, can be approximated
arbitrarily closely, and  also that the Bell
inequality can be violated `almost' maximally by selective, distant measurements
- even in the vacuum state. For a detailed discussion of the
philosophical import of these results reference may be made to \cite
{red5} and \cite{red6}. For a comprehensive discussion of the relation
between entanglement and correlation see \cite{clifty}.

\vspace {1cm}
{\bf Acknowledgement:} We are very grateful to David Malament for
proposing the root theorem to us, and fruitful discussion of our work.

\begin {appendix}
\section{Appendix}\label {app1}

{\bf\sc Proof of Lemma \ref{l1}}:
Choose $0<\varepsilon_1$ (without loss of generality
 $\varepsilon_1 <1$) and also $\|\psi\|=1$. Then, 
by Reeh and Schlieder $\|\psi-\tilde{C}_{\varepsilon_1}\Omega\|< \varepsilon_1$ for some operator
$\tilde{C}_{\varepsilon_1} \in {\cal R}(O_1)$.
Thus, 
\be
\varepsilon_1>\|\psi-\tilde{C}_{\varepsilon_1}\Omega\|\ge\left|\|\psi\|-
\|\tilde{C}_{\varepsilon_1}\Omega\|\right|=
\left|1-\|\tilde{C}_{\varepsilon_1}\Omega\|\right|\nonumber
\e
which implies that $\|\tilde{C}_{\varepsilon_1}\Omega\|>1-\varepsilon_1$. Now 
choose $C_{\varepsilon_2}$ such that $C_{\varepsilon_2}\Omega=\tilde{C}_{\epsilon_1}\Omega
/\|\tilde{C}_{\epsilon_1}\Omega\|$. So, 
\be
\|\psi-C_{\varepsilon_2}\Omega\|&=& \left\|\psi-\frac{\tilde{C}_{\varepsilon_1}\Omega}
{\|\tilde{C}_{\varepsilon_1}\Omega\|}\right\| =
\frac{1}{\|\tilde{C}_{\varepsilon_1}\Omega\|}\left\|\tilde{C}_{\varepsilon_1}\Omega-
\|\tilde{C}_{\varepsilon_1}\Omega\|\psi\right\| \nonumber\\ \vspace{5cm}
&=&\frac{1}{\|\tilde{C}_{\varepsilon_1}\Omega\|}\left\|\tilde{C}_{\varepsilon_1}\Omega-
\psi +\psi - 
\|\tilde{C}_{\varepsilon_1}\Omega\|\psi\right\| \nonumber\\
&\le&\frac{1}{\|\tilde{C}_{\varepsilon_1}\Omega\|}\left[\|\psi-\tilde{C}_{\varepsilon_1}\Omega\|
+\left\| \psi - 
\|\tilde{C}_{\varepsilon_1}\Omega\|\psi\right\|\right] \nonumber\\
&<& \frac{1}{1-\varepsilon_1}\left[\varepsilon_1+\left|1-\|\tilde{C}_{\varepsilon_1}\Omega
\|\right|\|\psi\|\right] < \frac{2\varepsilon_1}{1-\varepsilon_1}\stackrel
{\rm def.}{=:}\varepsilon_2.
\nonumber
\e
Clearly $\|C_{\varepsilon_2}\Omega\|=1$ and thus, lemma 1 is proved.\hfill
$\lhd$\\

\vspace{1cm}

{\bf\sc Proof of Lemma \ref{l2}:} The reader might verify that for all $\varepsilon_3>0$,
the operator $C_{\varepsilon_2}$ with $\varepsilon_2=-\frac{1}{2}+\sqrt{\frac{1}{4}+\frac
{\varepsilon_3}{\|A\|}}$ has the required property. However, it is instructive 
to follow the calculations that led to this value for $\varepsilon_2$. For
convenience set $\psi-C_{\varepsilon_2}\Omega=\mu$ such that $\|\mu\|<\varepsilon_2$.
We calculate 
\be
K&=& (\psi, A\psi)=(\mu+C_{\varepsilon_2}\Omega, A\mu + AC_{\varepsilon_2}\Omega)\nonumber\\
&\le& |(\mu, A\mu)|+|(C_{\varepsilon_2}\Omega, A\mu)|+(C_{\varepsilon_2}\Omega,
AC_{\varepsilon_2}\Omega)+ |(\mu, AC_{\varepsilon_2}\Omega)|\nonumber
\e
by the triangle inequality. Using the (Cauchy-)Schwarz(-Bunjakowski) inequality,
we find
\be
K&\le&\|\mu\|\|A\mu\| + \|C_{\varepsilon_2}\Omega\|\|A\mu\| + \langle A
\rangle_{C_{\varepsilon_2}\Omega} +\|A\mu\| \|C_{\varepsilon_2}\Omega\|\nonumber
\\
&\le& \|\mu\|^2\|A\| + \|C_{\varepsilon_2}\Omega\| \|\mu\| \|A\| + \|\mu\|
\|C_{\varepsilon_2}\Omega\| \|A\| + \langle A
\rangle_{C_{\varepsilon_2}\Omega}\nonumber\\
&<& (\varepsilon_2^2+ 2\varepsilon_2) \|A\| + \langle A\rangle_{C_{\varepsilon_2}\Omega}\nonumber
\e
Thus, $\langle A\rangle_{C_{\varepsilon_2}\Omega}>K-\varepsilon_3$, where
$\varepsilon_3= (\varepsilon_2^2 + 2\varepsilon_2)\|A\|$.
For the second inequality consider
\be
\langle A\rangle_{C_{\varepsilon_2}\Omega}-K&=&
-(\mu, A\mu)-(C_{\varepsilon_2}\Omega, A\mu)- (\mu, AC_{\varepsilon_2}\Omega)\nonumber\\
&\le&|(\mu, A\mu)|+|(C_{\varepsilon_2}\Omega, A\mu)|+|(\mu, AC_{\varepsilon_2}\Omega)|\nonumber\\
&<& (\varepsilon_2^2 + 2\varepsilon_2)\|A\|=\varepsilon_3\nonumber.
\e
 and thus, lemma \ref{l2} is proved.
\hfill $\lhd$\\

\vspace{1cm}

{\bf\sc Proof of Lemma \ref{l3}:} Clearly $\langle P_i \rangle_{\Omega}\neq 0\;
\forall i$, and hence $\langle \tilde{Q_1}'\rangle_{\Omega}\neq 0$. This follows
just from the fact that $\Omega$ is a separating vector as was shown above.
Now set $Q_1'=\tilde{Q_1}'/\langle \tilde{Q_1}'\rangle_{\Omega}$ in order to find
\be
\|Q_1-Q_1'\|&=&\left\|Q_1-\frac{\tilde{Q_1}'}{\langle \tilde{Q_1}'\rangle_{\Omega}}\right\|
=\frac{1}{\langle \tilde{Q_1}'\rangle_{\Omega}}\|\langle \tilde{Q_1}'\rangle_{\Omega}Q_1
-\tilde {Q_1}'\|\nonumber\\
&=& \frac{1}{\langle \tilde{Q_1}'\rangle_{\Omega}}\left\|Q_1\left(\langle \tilde{Q_1}'\rangle_{\Omega}
-\langle Q_1\rangle_{\Omega} \right)+ \langle Q_1\rangle_{\Omega}Q_1 - \tilde{Q_1}'\right\|\nonumber\\
&\le&\frac{1}{\langle \tilde{Q_1}'\rangle_{\Omega}}\left[\|Q_1\|\left|
\langle \tilde{Q_1}'\rangle_{\Omega}-\langle Q_1\rangle_{\Omega}\right|+
\|Q_1\|\left|\langle Q_1\rangle _{\Omega}-1\right| + \|Q_1 - \tilde{Q_1}'\|\right]\nonumber
\e
where we have used the triangle inequality and the fact that $Q_1$ is bounded. The
first term can be majorized as follows:
\be
\|Q_1\|\left|\langle \tilde{Q_1}'\rangle_{\Omega}-\langle Q_1\rangle_{\Omega}\right|
\le \|Q_1\| \|Q_1 -\tilde{Q}_1'\|<\|Q_1\|\tilde{\varepsilon}_4\nonumber
\e
The second term vanishes identically according to lemma \ref{l1}, and the third
term is smaller than $\tilde{\varepsilon}_4$. Hence
\be
\|Q_1-Q_1'\|<\frac{1}{\langle \tilde{Q_1}'\rangle_{\Omega}}\left[\|Q_1\|+1\right]
\tilde{\varepsilon}_4 \stackrel{\rm def.}{=:}\varepsilon_4\nonumber
\e
Thus, the lemma is proved as $\langle Q_1'\rangle_{\Omega}=1$, trivially.
\hfill$\lhd$\\

\vspace{1cm}

{\bf\sc Proof of Lemma \ref{l4}:} We know already from lemma \ref{l2} 
that for all $\varepsilon_3>0$
there is an operator $Q_1$ such that $K-\varepsilon_3<(\Omega, AQ_1\Omega)
< K+ \varepsilon_3$. Thus,
\be
K-\varepsilon_3&<&(\Omega, A(Q_1-Q_1'+Q_1')\Omega)=(A\Omega, (Q_1-Q_1')\Omega)
+(\Omega, AQ_1'\Omega)\nonumber \\
&\le& \|A\| \|Q_1-Q_1'\|+ \langle AQ_1'\rangle_{\Omega} < \|A\|\varepsilon_4
+\langle AQ_1'\rangle_{\Omega} \nonumber 
\e
and hence
\be
\langle AQ_1'\rangle_{\Omega}&>&K-\left(\varepsilon_3+ \|A\|\varepsilon_4\right) 
\stackrel{\rm def.}{=:}K-\varepsilon_5\nonumber.
\e
Similarly
\be
\langle A Q_1'\rangle_{\Omega}&=&(\Omega, AQ_1'\Omega)=(\Omega, A(Q_1'-Q_1+Q_1)\Omega)\nonumber\\
&=& (A\Omega, (Q_1'-Q_1)\Omega)+ (\Omega, AQ_1\Omega)\nonumber\\
&\le& \|A\| \|Q_1'-Q_1\| + \langle AQ_1\rangle_{\Omega}<\varepsilon_4\|A\|+
\langle AQ_1\rangle_{\Omega}\nonumber\\
&<& \varepsilon_4\|A\| + K + \varepsilon_3 = K + \varepsilon_5\nonumber
\e
which proves lemma \ref{l4} . \hfill$\lhd$\\
\end{appendix}


\addcontentsline {toc} {section} {References}
\begin{thebibliography}{99}
\renewcommand{\baselinestretch}{1.0}
\small
\bibitem {butter} J. Butterfield: {\em Vacuum Correlations and Outcome
Dependence in Algebraic Quantum Field Theory}, Annals of the New York Academy
of Sciences, Vol. 755 (1995), pp. 768-85.
\bibitem {clifty} R.K. Clifton, D.V. Feldman, M.L.G. Redhead,  and
A. Wilce: {\em Hyperentangled states}, {\ttfamily quant-ph/9711020}.
\bibitem {driessler} W. Driessler: {\em Comments on Lightlike Translations
and Applications in Relativistic Quantum Field Theory}, Commun. Math. Phys., Vol. 44
(1975), pp. 133-141.
\bibitem {epr} A. Einstein, B. Podolsky, N. Rosen: {\em Can quantum-mechanical
description of physical reality be considered complete}, Phys. Rev., Vol. 47
(1935), pp. 777-80. Reprinted in: J. A. Wheeler and W.H. Zurek (eds.):
{\em Quantum Theory and Measurement}, Princeton: Princeton University Press, 1983, 
pp. 138-41.
\bibitem {fred} K. Fredenhagen: {\em A Remark on the Cluster Theorem},
Commun. Math. Phys., Vol. 97 (1985), pp. 461-3
\bibitem {haag} R. Haag: {\em Local Quantum Physics}, corrected 2nd
Printing, Berlin/New York: Springer Verlag, 1993.
\bibitem {hkr} K.-E. Hellweg and K. Kraus: {\em
Operations and Measurements II}, Commun. Math. Phys., Vol 16 (1970), 142-7.
\bibitem {horuzhy} S. S. Horuzhy: {\em Introduction to Algebraic Quantum 
Field Theory}, Dordrecht: Kluwer Academic Publishers, 1990.
\bibitem {landau1} L.J. Landau: {\em On the Violation of Bell's
Inequality in Quantum Theory}, Phys. Lett. A, Vol. 120, No. 2 (1987), pp. 54-6.
\bibitem {landau2} --- : {\em On the Non-Classical Structure of the Vacuum}, 
Phys. Lett. A, Vol. 123, No. 3 (1987), pp.115-8.
\bibitem {licht} A. L. Licht: {\em Local States}, J. Math. Phys., Vol. 7, 
No. 7 (1966), pp. 1656-69.
\bibitem {mall} D. Malament, private communication.
\bibitem {red4} M.G.L. Redhead: {\em Incompleteness, Nonlocality, and Realism},
Oxford: Oxford University Press, 1987. 
\bibitem {red3}---: {\em More Ado about Nothing}, Found. Phys., 
Vol. 25, No. 1 (1995), pp. 123-137.
\bibitem {red5}---: {\em The Vacuum in Relativistic Quantum Field
Theory}, in D. Hull, M. Forbes and R.M. Burian (eds.): {\em
Proceedings of the Biennial Meeting of the Philosophy of Science
Association}, Vol. 2, East Lansing: PSA, 1995, pp. 77-87.
\bibitem {red6}--- and P. La Rivi\`ere: {\em The Relativistic EPR
Argument}, in R. Cohen, M. Horne and J. Stachel (eds.): {\em
Potentiality, Entanglement and Passion-at-a-Distance: Quantum
Mechanical Studies for Abner Shimony}, Vol. 2, Dordrecht: Kluwer,
1997, pp. 207-215.
\bibitem {reed} M. Reed and B. Simon: {\em Methods of modern mathematical
physics, Vol. 1: Functional Analysis}, revised and enlarged edition, San Diego:
Academic Press, 1980.
\bibitem {reeh} H. Reeh and S. Schlieder: {\em Bemerkungen zur Unit\"ar\"aquivalenz
von Lorentzinvarianten Feldern}, Il Nuovo Cimento, Vol. 22, No. 5 (1961), pp. 
1051-68.
\bibitem {s1} S. J. Summers: {\em On the independence of local algebras
in quantum field theory}, Rev. Math. Phys., Vol. 2, No. 2 (1990), pp. 201-47.
\bibitem {summ1} Summers, S. J. and Werner, R.: {\em The Vacuum Violates 
Bell's Inequalities}, Phys. Lett. A, Vol. 110, No. 5 (1985), pp. 257-9.
\bibitem {summ2} --- : {\em Bell's inequalities and quantum field
theory, I. General setting}, J. Math. Phys, Vol. 28, No. 10 (1987), pp. 2440-7.
\bibitem {summ3} --- : {\em Bell's inequalities and quantum field
theory, II. Bell's inequalities are maximally violated in the vacuum}, 
J. Math. Phys, Vol. 28, No. 10 (1987), pp. 2448-56.
\bibitem {summ4} --- : {\em Maximal Violation of Bell's 
Inequalities is Generic in Quantum Field Theory}, Commun. Math. Phys., 
Vol. 1, No. 10 (1987), pp. 247-59.
\end{thebibliography}
\end{document}